\documentclass[12pt]{article}

\input epsf.tex 

\def\np#1#2#3{Nucl. Phys. B{#1} (#2) #3}
\def\pl#1#2#3{Phys. Lett. {#1}B (#2) #3}
\def\prl#1#2#3{Phys. Rev. Lett. {#1} (#2) #3}
\def\physrev#1#2#3{Phys. Rev. {D#1} (#2) #3}

\def\ptp#1#2#3{Prog. Theor. Phys. #1 (#2) #3}

\def\ltap{\ \raise.3ex\hbox{$<$\kern-.75em\lower1ex\hbox{$\sim$}}\ } 
\def\gtap{\ \raise.3ex\hbox{$>$\kern-.75em\lower1ex\hbox{$\sim$}}\ }

 \def\gev{\,{\rm GeV}}
\def\be{\begin{equation}} \def\ee{\end{equation}}
 
\def\abs#1{\left| #1 \right|} 

\newcommand{\drawsquare}[2]{\hbox{% 
\rule{#2pt}{#1pt}\hskip-#2pt% left vertical
\rule{#1pt}{#2pt}\hskip-#1pt% lower horizontal
\rule[#1pt]{#1pt}{#2pt}}\rule[#1pt]{#2pt}{#2pt}\hskip-#2pt% upper horizontal
\rule{#2pt}{#1pt}}% right vertical

\newcommand{\Yfund}{\raisebox{-.5pt}{\drawsquare{6.5}{0.4}}}
\newcommand{\Yasymm}{\raisebox{-3.5pt}{\drawsquare{6.5}{0.4}}\hskip-6.9pt
\raisebox{3pt}{\drawsquare{6.5}{0.4}}}
\newcommand{\bYfund}{\overline{\Yfund}}

\begin{document}

\pagestyle{plain}
\setcounter{page}{1}
\newcounter{bean}
\baselineskip16pt
\begin{titlepage}
\begin{flushright}
PUPT-1721\\
hep-ph/9709383
\end{flushright}

\vspace{7 mm}

\begin{center}
{\large \bf New Models of Gauge Mediated }

\vspace{2mm}
{\large \bf Dynamical Supersymmetry Breaking}
\end{center}
\vspace{10 mm}
\begin{center}
{
Yuri Shirman\footnote{\tt email:
yuri@feynman.princeton.edu}\\
}
\vspace{3mm}
{\em Joseph Henry Laboratories\\
Princeton University\\
Princeton, New Jersey 08544}
\end{center}
\vspace{7mm}
\begin{abstract}
\noindent
We propose a simple class of nonrenormalizable models of gauge mediated
dynamical supersymmetry breaking. 
The models do not have gauge singlet fields.
The Standard Model gauge group is embedded in 
the global symmetry of the SUSY
breaking sector.
At the renormalizable level the models possess
a set of classical flat directions. Only one of those flat directions is
unlifted by quantum effects, and requires nonrenormalizable term 
to stabilize the potential for the corresponding modulus. Large
vacuum expectation value of this modulus at the minimum of the potential
generates mass terms for the messenger fields. There are no light
messengers, thus this class of models evades difficulties encountered
in earlier constructions using nonrenormalizable models.
\end{abstract}
 
\vspace{7mm}
\begin{flushleft}
September 1997

\end{flushleft}
\end{titlepage}

\newpage

%\section{Introduction}

Gauge mediated supersymmetry breaking (GMSB) \cite{original} offers an
attractive solution to the problem of the flavor 
changing neutral currents.
While the original explicit models of GMSB \cite{dn} $-$ \cite{ dnns} are
phenomenologically viable, they are quite complicated.  
This is because the most
elegant idea \cite{adsdsb} of identifying Standard Model 
gauge group with the
global symmetry of the supersymmetry breaking sector leads to 
the large increase in
the number of fields with the Standard Model gauge 
quantum numbers and as a
result QCD becomes non-asymptotically 
free and hits its Landau pole just a few
decades above the weak scale.  
One possible solution to this problem suggested in \cite{dn}
is to isolate the dynamical supersymmetry breaking 
(DSB) sector of the model
from the Standard Model sector. This solution while leading to
realistic models is not very appealing.

Recently a lot of effort has been devoted 
to simplifying the structure of the
GMSB models \cite{ptdirect}$-$\cite{otherdirect}.  
Here we will follow an approach which attempts to identify
the Standard Model with the (weakly gauged) subgroup of the
global symmetry of the DSB sector. The asymptotic freedom
problem mentioned above can be solved by making messenger fields
very heavy so that 
they do not significantly affect running of
the gauge couplings up to the 
unification scale and do not spoil perturbative
gauge coupling unification. To generate
large masses for the messengers one couples
them to a modulus which acquires large vacuum expectation value 
(vev) at the minimum of the
potential. Several classes of models have been constructed along
these lines \cite{ptdirect}$-$\cite{luty}.
We will briefly review two of them here.

In the class of models suggested by Poppitz and Trivedi 
\cite{ptdirect} the 
modulus parameterizes a D-flat direction which is only lifted
by a non-renormalizable operator, and, therefore, 
obtains a large vev at the
minimum of the potential.  The serious problem 
encountered by Poppitz and
Trivedi (which was also present in the analogous models of ref.
\cite{mrmah}) is that there are light messenger 
fields with significant
soft SUSY breaking scalar masses and positive supertrace.  
This leads to the negative mass squared
for the squarks and sleptons through the two-loop RGE 
evolution \cite{mrmah, negative}

Another mechanism \cite{murayama, ddgr} to generate large vev 
for the modulus is through a modification
of the quantum moduli space models \cite{IT, japan} 
in which large vev is generated by the inverted hierarchy mechanism
\cite{inverted}.  
The most elegant models \cite{murayama, ddgr} 
constructed along these lines suffer from the following
problem (see \cite{simple} for the possible resolution). 
Since the Standard Model gauge group is
identified with an unbroken diagonal subgroup of 
the product gauge group of the
microscopic theory, models possess gauge messengers.  As was 
shown in \cite{gr} this leads to the significant 
negative contribution to the superpartner masses.

Here we suggest a class of models which
 circumvent the difficulties mentioned
above.  This class of models can be thought 
of as a hybrid between the two
approaches.  In our models there will 
be a set of classical flat directions
unlifted by the tree level superpotential 
at the renormalizable level.  All but
one of them will be lifted quantum mechanically.  
The quantum effects will lead to
generation of a run-away scalar potential along remaining 
classical flat direction as in the models of refs. \cite{ptdirect, mrmah}.
The scalar potential will be stabilized
by the nonrenormalizable operator in the tree level superpotential, 
ensuring that the
corresponding modulus will acquire a large vev and generate large
masses for all messengers. Both modulus and the messenger 
fields originate in a sector very similar to the 
$SU(5)^3$ model of ref. \cite{ddgr}. In our case, however,
the Standard Model matter fields and the modulus do not carry
quantum numbers under the same gauge groups and thus gauge messengers
do not appear.

%\section{An SU(5)$^3$ Model}

The simplest model in this class is
based on $SU(5)_1 \times SU(5)_2 \times SU(5)_G$ 
gauge group with a matter
content given in the table 1.  
The Standard Model gauge group will
be embedded
in $SU(5)_G$, but in our discussion of the
supersymmetry breaking we will treat $SU(5)_G$ as a global 
symmetry.  As usual we will work 
in terms of the complete 
GUT multiplets although this is not essential.  We will choose to
write the tree level superpotential in the form 
\be\label{wtree} W=XQ\bar Q +
{1\over M^2_{Pl}} X^5 + \cdots 
\ee 
We could also have added other nonrenormalizable
terms, such as $Q^5$ and $\bar Q^5$. 
However, we will be interested in the
dynamics of the model for large $X$ where such terms are negligible.  In
addition one could impose symmetries to exclude these terms.
\begin{table} 
\begin{center}
%\vskip .2in 
\begin{tabular}{|l|c|c|c|} 
\hline \hfil &\hfil SU(5)$_1$ \hfil &\hfil SU(5)$_2$ \hfil  &\hfil 
SU(5)$_G$ \hfil \\ \hline &&& \\ [-8pt] 
$A$ &$\Yasymm$ &1 &1 \\[2pt] 
$\bar F$ &$\bYfund$ &1 &1 \\ \hline &&&  \\ [-8pt] 
$X$ &$\bYfund$ &$\Yfund$ &1 \\ 
$\bar Q$ &$\Yfund$ &1 &$\bYfund$ \\ 
$Q$& 1 &$\bYfund$ &$\Yfund$ \\ \hline 
\end{tabular} 
%\vskip .2 in 
\caption{Quantum numbers
of chiral superfields in SU(5)$^3$ model} 
%\vskip .2 in 
\end{center}
\end{table} 
 
Let us start by commenting on
the matter content of the model.  It clearly consists of two distinct
sectors.  The first sector 
contains antisymmetric tensor $A$ and antifundamental
$\bar F$ charged under $SU(5)_1$ group only.  
This sector has the matter content
of the
supersymmetry
breaking $SU(5)$ model \cite{adsdsb}.  
Our goal is to construct the full model in such 
a way that the low-energy
effective theory describing SUSY breaking would 
contain $A$ and $\bar F$ as the
light fields with the addition of the modulus from the second 
sector\footnote{This low energy matter content is analogous
to the models of ref. \cite{luty}. In \cite{luty} , however,
the strong $SU(5)$ dynamics served to stabilize
modulus, while in our models it will push modulus to large
vev.}.  We
will, therefore, loosely refer to this sector as a DSB sector.  The second
sector is a model discussed in \cite{pts} with an $SU(N)^2$ gauge group
and $N_f=N$ flavors for each group.  
This sector has a run-away
direction\footnote{This is exactly the direction we are interested in
and it will persist in the full model at the renormalizable level. 
The dynamics along this direction will, however, be somewhat modified
by the presence of the additional fields in the DSB sector.}
parametrized by the vev of the light modulus $v=(\det X)^{1/5}$.  All
other classical flat directions of 
this sector do not lead to supersymmetric
vacua.    
This sector will provide both messenger 
fields and the light modulus with
non-vanishing vacuum expectation value for 
both the scalar and auxiliary
component and we will refer to it as a messenger sector.

At the renormalizable level the 
superpotential (\ref{wtree}) possesses a set of
classical flat directions parametrized by the vev's of the gauge invariant
polynomials 
$S=v^5=\det (X)$, $B=\det (Q)$, $\bar B = \det(\bar Q)$, $P=A^2 \bar Q$,
$N=A \bar Q^3$ and $M=\bar F \bar Q$.  
As mentioned above we are interested in the dynamics for
large $v$.  In such a case all 
components of the $Q$ and $\bar Q$ become heavy.
One, therefore, can hope that all 
gauge invariant polynomials involving $Q$
or $\bar Q$ will have vanishing vev's.  
Still, let us show that all
classical flat directions but $v$ are 
lifted quantum mechanically.
\begin{description} 
\item{1.}  Along the $B$ direction the $SU(5)_2$ is
completely broken while $SU(5)_1$ remains unbroken.  
All matter fields but $A$
and $\bar F$ become heavy.  
The low energy dynamics therefore breaks
supersymmetry for every fixed value of $B$.  
By matching scales of
microscopic and low-energy theories, we find the potential for $B$
\be\label{norunone} 
V \sim \Lambda_L^4 \sim \left( B \Lambda_1^8
\right)^{4/13} 
\ee 
Clearly this stabilizes classical flat direction.  
\item{2.}
Along the $\bar B$ direction the gauge group $SU(5)_1$ is 
completely broken, while
$SU(5)_2$ remains unbroken.  There are no light matter 
fields charged under the
unbroken gauge group.  Gaugino condensation 
generates the superpotential.
Using the scale matching conditions we can 
find the superpotential for $\bar B$:
\be\label{noruntwo} 
W=\Lambda_L^3=\bar B^{1/5} \Lambda_2^2 
\ee 
which leads to a
$\bar B$-independent non-vanishing 
potential\footnote{Remember that K\"ahler
potential is nearly canonical 
in terms of elementary quark superfields for large
$\bar B$.}.  
At one loop a potential for $\bar B$ 
is generated and theory is stabilized
near the origin \cite{murayama, ddgr, flat} 
(for certain range of parameters there is a
local SUSY breaking vacuum for large but finite $\bar B$   due to the
contributions of the dynamics in the broken $SU(5)_1$ group \cite{
murayama, ddgr}).  
\item{3.}  Along $P$ and $N$ flat directions 
unbroken gauge group is again $SU(5)_2$.  In both cases
there are 5 flavors transforming under 
the strong $SU(5)_2$ group.  They are
coupled to $\bar Q$ fields\footnote{Due 
to the large vev of the component(s) of $\bar Q$ one of
the flavors is heavy along $P$  direction, and three flavors
are heavy along $N$ direction.  Thus we
could have considered effective $SU(5)$ with $N_f=4$ (or $N_f=2$)
and modulus-dependent
scale.  This would lead us to the same conclusions.} 
which are singlets of the $SU(5)_2$.  As is well known all
classical flat directions involving 
gauge singlet fields are lifted quantum
mechanically in such a case.  
\item{4.} $M$ direction is potentially the most dangerous. Along this
direction $SU(5)_1$ is only broken to $SU(4)_1$ subgroup with a scale
inversely proportional to a power of the vev. If the nonperturbative
superpotential were generated by the strong $SU(4)_1$ 
dynamics, the interference
effects could potentially 
lead to the restoration of the supersymmetry. Fortunately,
there are 5 flavors of fundamentals fields and antisymmetric tensor
transforming under 
the effective $SU(4)_1$ group, and no superpotential can be generated. 
Therefore, $SU(4)_1$ dynamics can be neglected
for large vev. Repeating the previous analysis of the $SU(5)_2$ dynamics
we conclude that $M$ is lifted.
\end{description}

It is also easy to show that there is no SUSY minimum near the
origin of the moduli space. Consider a limit 
$\Lambda_2 \gg \Lambda_1$. Below the scale $\Lambda_2$
the renormalizable Yukawa coupling in the superpotential
turns into the mass term for $SU(5)_2$ mesons and $\bar Q$. 
At low energies the effective description is SUSY breaking
$SU(5)$ model with the scale 
$\Lambda_L^{13}=\Lambda_1^8 \Lambda_2^5$. 
As a result there is no SUSY vacuum near the origin.
Light spectrum also contains
baryons
$S=X^5$ and $B$ which are singlets under the low energy gauge group.
For $S, B \ll \Lambda_2$ the K\"ahler
potential is nearly canonical in terms of baryonic variables.
The $B$ directions is lifted according to eqn. (\ref{norunone}).
Due to the quantum modified constrain in $SU(5)_2$ gauge group
this leads to $S$ acquiring vev - thus leading the model towards
the vacuum of interest.

Having established that all unwanted classical flat directions are lifted
quantum mechanically and that there is no SUSY minimum 
near the origin of the moduli space, we are ready to consider the
effective theory for large $v$.
In this case the gauge group is broken to the diagonal $SU(5)_L$.  
In the effective
theory only $A$, $\bar F$, and $v$ remain light. 
 For every fixed value of $v$
the potential is nonvanishing.  
The model is noncalculable and we can only give estimates of
the vacuum energy and other parameters at the minimum of 
the scalar potential\footnote{Note that despite this fact our 
model is quite predictive, since
holomorphic SUSY breaking contributions to messenger masses dominate.
The Standard Model superpartner spectrum can 
easily be calculated in terms of
$\Lambda_{SUSY}=F_v /v$ and $\mu$, assuming that dynamics generating
$\mu$ term is not connected with supersymmetry breaking.}.
Using the scale matching conditions we find
\be\label{runnow} 
V \sim \Lambda_L^4 \sim \left({\Lambda_1^8 \Lambda_2^{10}
\over v^5}\right)^{4/13} 
\ee 
This leads to run-away behavior.  When we turn on
nonrenormalizable coupling,  the scalar potential is stabilized.  
Vacuum energy will be
determined by the balance between the potential in (\ref{runnow}) and
$\abs{F_v}^2 \sim \abs{{v^4\over M_{Pl}^2} }^2$. At the minimum  
\be\label{minimum} 
v \sim
\left( M_{Pl}^{13} \Lambda^{18}\right ) ^{1\over 31} ~~~ V \sim
\left({\Lambda^{144} \over M_{Pl}^{20}} \right)^{1\over 31} ~~~ F_v \sim
\left({\Lambda^{72} \over M_{Pl}^{10}} \right)^{1\over 31} ~~~ 
{F_v\over v} \sim
\left ({\Lambda^{54} \over M_{Pl}^{23}} \right)^{1\over 31} 
\ee 
where we used notation
$\Lambda^9=\Lambda_1^4\Lambda_2^5$.

Upon identifying an $SU(3)\times SU(2) \times U(1)$ 
subgroup of the global
$SU(5)_G$ symmetry with the 
Standard Model the heavy fields $Q$ and $\bar Q$
serve as messengers of the supersymmetry breaking.  If we require 
that the scale of the supersymmetry breaking
breaking in the Standard Model sector be 
$\Lambda_{SUSY} = {F_v \over v} \sim 10^4 \gev$ we find that 
\be\label{numbers} 
\Lambda \sim {\rm few~} \times
10^{10}~ \gev ~~~ v \sim {\rm few~} \times 10^{13}~ \gev ~~~ 
\sqrt{F} \sim {\rm few} \times 10^8~ \gev 
\ee

It is easy to find several generalizations of 
the $SU(5)^3$ model described above.
An almost trivial modification involves 
interchange of fundamental and antifundamental
fields in the messenger sector. 
The set of mixed flat directions in such a model
is different. Still only the $v$ direction is not stabilized at the
renormalizable level, and the dynamics along 
this direction is  the same as discussed above.
More generally one can use a different DSB sector.  
Any DSB model without classical
flat directions and a gauge group with $SU(5)$ factor can be a candidate.  
Such a
modification clearly does not change our discussion of 
the $B$ and $\bar B$ classical flat directions.  
One should carefully check the stabilization of the
potential along mixed flat directions involving vev's 
of the fields of both sectors.  In particular, it is possible that
superpotential is generated in the unbroken subgroup of the DSB sector.
If this is the case, it is necessary to verify that there are no
interference effects leading to the supersymmetry 
restoration\footnote{Even when this happens 
there generically will be a local SUSY 
breaking minimum for large $v$.}.
\begin{table}
\begin{center} 
%\vskip .2in 
\begin{tabular}{|l|c|c|c|c|c|} 
\hline \hfil &\hfil
U(1) \hfil & \hfil SU(2) \hfil &\hfil SU(5)$_1$ \hfil &\hfil SU(5)$_2$ 
\hfil&\hfil
SU(5)$_G$ \hfil \\ 
\hline &&&&& \\ [-8pt] 
$A$ &4 &1 &$\Yasymm$ &1 &1 \\[2pt] $F$
&-3 &$\Yfund$ &$\Yfund$ &1 &1 \\ 
$\bar F_i, i=1\ldots 3$ &-2 &1 &$\bYfund$ &1 &1\\ 
$\phi_i, i=1\ldots 3$ &5 &$\Yfund$ &1 &1 &1 \\ 
$S$ &-10 &1 &1 &1 &1 \\
\hline &&& \\ [-14pt] 
$X$ &0 &1 &$\Yfund$ &$\bYfund$ &1 \\ 
$Q$ &0 &1 &$\bYfund$ &1 &$\Yfund$\\ 
$\bar Q$ &0 &1 &1 &$\Yfund$ &$\bYfund$ \\ \hline 
\end{tabular}
%\vskip .2 in 
\caption{Quantum numbers of chiral superfields in a
calculable model} 
%\vskip .2 in 
\end{center}
\end{table}

As an example consider a model based on the 
$U(1) \times SU(2)\times SU(5)_1\times SU(5)_2
\times SU(5)_G$ group with matter content 
as given in table 2 and tree level
superpotential 
\be\label{calculable} 
W= \gamma A \bar F_1 \bar F_2 + \eta S
\phi_1 \phi_2 + \delta_i F \bar F_i \phi_i + \lambda X Q \bar Q 
+{1\over M_{Pl}^2} X^5
\ee 
The DSB
sector of this model is described in ref. \cite{dnns}.  
While the dynamics along mixed flat directions is
much more complicated, it does not lead to the runaway behavior.
For large $v$ the effective description
is the $SU(5) \times SU(2) \times U(1)$ model of ref. \cite{dnns}
with the strong coupling scale given by 
$\Lambda^{11} = \Lambda_1^6 \Lambda_2^{10} / v^5$.  
Thus we have an example
in which low energy description is given by  a calculable model.

%\bigskip {\it $F$-term for the modulus}

For our models to lead to 
a realistic spectrum of the superpartner masses it is
important for both the scalar component and 
$F$-term of the modulus $v$ to have
a non-vanishing expectation value.  
In the $SU(5)^3$ model discussed above one
can not calculate an $F$ term for the light modulus.  
We could only establish
the order of magnitude of this term on dimensional grounds.  
While there are no
symmetry reasons in this model for the $F$-term to vanish, it would be
satisfying to check the assertion in a calculable model.  
Our second example, 
an $U(1) \times SU(2)\times SU(5)_1 \times SU(5)_2$ model, 
is calculable and presents such a
possibility.  Instead of minimizing potential 
of this model we will work with a
toy example\footnote{We thank Yael Shadmi for suggesting this example.} 
based on the $SU(2)\times SU(3)_1\times SU(3)_2 \times SU(3)_G$ 
model\footnote{This
model has only $SU(3)_G$ global symmetry, and can not be used for model
building.}  with a matter content given in the table 3.  
For large $v$ the effective
description is the $3-2$ model of 
Affleck-Dine-Seiberg \cite{adsdsb} with a modulus
dependent scale and the effective superpotential 
(after integrating out heavy fields) 
\be\label{threetwo} 
W={\Lambda^{10} \over v^3 \det(q \bar q)} +
\lambda_1 q \ell \bar d + \lambda_2 v^3 
\ee 
where
$\Lambda^{10}=\Lambda_1^4\Lambda_2^6$, $q= (\bar u, \bar d)$,
and $\lambda_2$ is small.  
We find that
this model breaks SUSY and at the minimum 
\be\label{Fterm} 
F_v=0.3
\lambda_1^{4/15} \lambda_2^{8/15} \Lambda^2 
\ee 
\begin{table}
\begin{center} 
%\vskip .2in 
\begin{tabular}{|l|c|c|c|c|} 
\hline \hfil &\hfil
SU(2) \hfil &\hfil SU(3)$_1$ \hfil &\hfil SU(3)$_2$ \hfil&
\hfil SU(3)$_G$ \hfil \\ 
\hline 
&&& \\ [-14pt] 
$q$ &2 &3 &1 &1 \\ 
$\bar u$ &1 &$\bar 3$ &1 &1 \\ 
$\bar d$
&1 &$\bar 3$ &1 &1 \\ 
$\ell$ &2 &1 &1 &1 \\ \hline 
&&& \\ [-14pt] 
$X$ &1 &3 &$\bar 3$ &1 \\ 
$Q$&1 & $\bar 3$ &1 &3 \\ 
$\bar Q$&1 &1 &3 &$\bar 3$ \\ \hline
\end{tabular} 
%\vskip .2 in Table 3.  
\caption{ Quantum numbers of chiral superfields
in toy model} 
%\vskip .2 in
\end{center} 
\end{table}
Therefore, we have established that
light modulus has non-vanishing $F$-term as desired for model-building.

Finally, let us comment on the $\mu$-problem. It is as severe in our
models as in the most other GMSB models (see,  however, ref. \cite{simple}).
One could use a horizontal symmetry as suggested in \cite{variations}
to generate $\mu$-term while a small (order $\alpha_2$) $B$-term
would be generated at the two loop level.

To summarize, we have presented here
a simple class of models with gauge mediated 
dynamical supersymmetry breaking. 
While implementing the idea of direct gauge mediation our models
avoid some of the difficulties encountered in earlier attempts to realize
this approach. In particular all messenger fields are heavy and
there are no gauge messengers. Our models are completely 
chiral and do not contain gauge singlets, although gauge singlet 
fields may have to be introduced to generate $\mu$-term of the appropriate
order of magnitude. Some of our models are calculable, but even noncalculable
$SU(5)^3$ model is quite predictive.

\section*{ Acknowledgements} I would like to thank Michael Dine,
 Ann Nelson and Yael Shadmi for useful discussions and comments.
I am also grateful to Theory Group at Fermilab, where this work
has begun, for hospitality. This work was supported in part by NSF
grant PHY-9157482 and James S. McDonnell Foundation grant No. 91-48.


\begin{thebibliography}{40}

\bibitem{original}M. Dine, W. Fischler, M. Srednicki, \np{189}{575}{1981};
S. Dimopoulos, S. Raby, \np{192}{1981}{353};
M. Dine, W. Fischler, \pl{110}{1982}{227}; M. Dine, M. Srednicki, 
\np{202}{1982}{238}; M. Dine, W. Fischler, \np{204}{1982}{346};
L. Alvarez-Gaum\'e, M. Claudson, M. Wise, \np{207}{1982}{96};
C. Nappi, B. Ovrut, \pl{113}{1982}{175}; S. Dimopoulos, S. Raby, 
\np{219}{1983}{479}.
\bibitem{dn}A. Nelson, M. Dine, \physrev{48}{1993}{1277}, hep-ph/9303230.
\bibitem{dns} M. Dine, A. E. Nelson and Y. Shirman, \physrev{51}{1995}{1362},
hep-ph/9408384.
\bibitem{dnns} M. Dine, A. Nelson, Y. Nir, Y. Shirman,
\physrev{53}{1996}{2658}, hep-ph/9507378.
\bibitem{adsdsb} I. Affleck, M. Dine, N. Seiberg, \np{256}{1985}{557}.
\bibitem{ptdirect}E. Poppitz, S. Trivedi, \physrev{55}{5508}{1997}, 
hep-ph/9609529.
\bibitem{mrmah}N. Arkani-Hamed, J. March-Russel, H. Murayama, hep-ph/9701286.
\bibitem{murayama}H. Murayama, \prl{79}{18}{1997}, hep-ph/9705271.
\bibitem{ddgr}S. Dimopoulos, G. Dvali, G. Giudice, R. Rattazzi,
hep-ph/9705307.
\bibitem{simple}S. Dimopoulos, G. Dvali, R. Rattazzi, hep-ph/9707537.
\bibitem{luty}M. Luty, hep-ph/9706554.
\bibitem{otherdirect}
T. Hotta, K.-I. Izawa, T. Yanagida, \physrev{55}{415}{1997}, hep-ph/9606203; 
N. Haba, N. Maru, T. Matsuoka, \np{497}{31}{1997}, hep-ph9612468; 
L. Randall, \np{495}{37}{1997}, hep-ph/9612246;
Y. Shadmi, \pl{405}{99}{1997}, hep-ph/9703312;
N. Haba, N. Maru, T. Matsuoka, \physrev{56}{4207}{1997}, hep-ph/9703250;
C. Csaki, L. Randall, W. Skiba, hep-ph/9707386.
\bibitem{negative}E. Poppitz, S. Trivedi, \pl{401}{38}{1997}, hep-ph/9703246.
\bibitem{IT} K. Intriligator, S. Thomas,  \np{473}{1996}{121}, hep-ph/9603158.
\bibitem{japan}K.-I. Izawa, T. Yanagida, \ptp{95}{1996}{829},
hep-ph/9602180.
\bibitem{inverted}E. Witten, \pl{105}{267}{1981}. 
\bibitem{gr}G. Giudice, R. Rattazzi, hep-ph/9706540.
\bibitem{pts}E. Poppitz, Y. Shadmi, S. Trivedi, \pl{388}{561}{1996},
hep-th/9606184.
\bibitem{flat}Y. Shirman, \pl{389}{1996}{287}, hep-th/9608147.
\bibitem{variations}M. Dine, Y. Nir, Y. Shirman, \physrev{55}{1997}{1501},
hep-ph/9607397.
\end{thebibliography}
\end{document}